\documentclass[a4paper,11pt]{article}
\pdfoutput=1 

\usepackage{jcappub} 

\renewcommand{\v}[1]{\ensuremath{\mathbf{#1}}} 
\newcommand{\gv}[1]{\ensuremath{\mbox{\boldmath$ #1 $}}} 
\newcommand{\avg}[1]{\left< #1 \right>} 
\renewcommand{\d}[2]{\frac{d #1}{d #2}} 
\newcommand{\pd}[2]{\frac{\partial #1}{\partial #2}} 
 
\newcommand{\grad}[1]{\gv{\nabla} #1} 
\let\baraccent=\= 
\renewcommand{\=}[1]{\stackrel{#1}{=}} 

\newcommand{\p}{^{\prime} }

\title{\boldmath Multiwavelength Analysis of Dark Matter Annihilation and RX-DMFIT}

\author[a,1]{A. McDaniel,\note{Corresponding author.}}
\author[a, b]{T. Jeltema,}
\author[a,b]{S. Profumo,}
\author[c]{E. Storm}

\affiliation[a]{Department of Physics, University of California, \\1156 High St. Santa Cruz, CA, 95064, USA}
\affiliation[b]{Santa Cruz Institute for Particle Physics, University of California, \\1156 High St. Santa Cruz, CA, 95064, USA}
\affiliation[c]{GRAPPA, Institute of Physics, Universiteit van Amsterdam\\ Science Park 904, 1098XH Amsterdam, The Netherlands}

\emailAdd{alexmcdaniel@ucsc.edu}
\emailAdd{tesla@ucsc.edu}
\emailAdd{profumo@ucsc.edu}
\emailAdd{e.m.storm@uva.nl}

\abstract{
Dark matter (DM) particles are predicted by several well motivated models to yield Standard Model particles through self-annihilation that can potentially be detected by astrophysical observations. In particular, the production of charged particles from DM annihilation in astrophysical systems that contain magnetic fields yields radio emission through synchrotron radiation and X-ray emission through inverse Compton scattering of ambient photons. We introduce RX-DMFIT, a tool used for calculating the expected secondary emission from DM annihilation. RX-DMFIT includes a wide range of customizable astrophysical and particle parameters and incorporates important astrophysics including the diffusion of charged particles, relevant radiative energy losses, and magnetic field modelling. We demonstrate the use and versatility of RX-DMFIT by analyzing the potential radio and X-ray signals for a variety of DM particle models and astrophysical environments including galaxy clusters, dwarf spheroidal galaxies and normal galaxies. We then apply RX-DMFIT to a concrete example using Segue I radio data to place constraints for a range of assumed DM annihilation channels. For WIMP models with $M_{\chi} \leq 100$ GeV and assuming weak diffusion, we find that the the leptonic $\mu^+\mu^-$ and $\tau^+\tau^-$ final states provide the strongest constraints, placing limits on the DM particle cross-section well below the thermal relic cross-section, while even for the $b\bar{b}$ channel we find limits close to the thermal relic cross-section. Our analysis shows that radio emission provides a highly competitive avenue for dark matter searches.
}

\begin{document}

\let\jnl@style=\rm
\def\reff@jnl#1{{\rm#1}} 

\def\araa{\reff@jnl{ARA\&A}}             
\def\aj{\reff@jnl{AJ}}                   
\def\apj{\reff@jnl{ApJ}}                 
\def\apjl{\reff@jnl{ApJ}}                
\def\apjs{\reff@jnl{ApJS}}               
\def\apss{\reff@jnl{Ap\&SS}}             
\def\aap{\reff@jnl{A\&A}}                
\def\aapr{\reff@jnl{A\&A~Rev.}}          
\def\aaps{\reff@jnl{A\&AS}}              
\def\baas{\ref@jnl{BAAS}}               
\def\jcap{\reff@jnl{J. Cosmology Astropart. Phys.}}
\def\jrasc{\reff@jnl{JRASC}}             
\def\memras{\reff@jnl{MmRAS}}            
\def\mnras{\reff@jnl{MNRAS}}             
\def\na{\reff@jnl{New A}}                
\def\nar{\reff@jnl{New A Rev.}}          
\def\pra{\reff@jnl{Phys.~Rev.~A}}        
\def\prb{\reff@jnl{Phys.~Rev.~B}}        
\def\prc{\reff@jnl{Phys.~Rev.~C}}        
\def\prd{\reff@jnl{Phys.~Rev.~D}}        
\def\pre{\reff@jnl{Phys.~Rev.~E}}        
\def\prl{\reff@jnl{Phys.~Rev.~Lett.}}    
\def\pasa{\reff@jnl{PASA}}               
\def\pasp{\reff@jnl{PASP}}               
\def\pasj{\reff@jnl{PASJ}}               
\def\qjras{\reff@jnl{QJRAS}}             
\def\physrep{\reff@jnl{Phys.~Rep.}}   

\let\astap=\aap
\let\apjlett=\apjl
\let\apjsupp=\apjs
\let\applopt=\ao
\maketitle
\flushbottom
\section{Introduction}
\label{sec:intro}
\subsection{Background and Motivation}

While studies of weak gravitational lensing, galaxy rotation curves, and angular anisotropies in the Cosmic Microwave Background (CMB) have shown the existence of dark matter that comprises roughly 25\% of the mass-energy of the universe \cite{Bergstrom}, the fundamental nature of dark matter has yet to be understood. Many well motivated particle theories suggest that a plausible dark matter candidate is a weakly-interacting massive particle (WIMP) \cite{Jungman, BertoneParticle}. Proposed dark matter WIMP models can undergo self-annihilation yielding standard model particles such as quarks, leptons, and bosons, which can then decay into charged particles such as electrons and positrons. The presence of these particles in astrophysical systems leads to unique signatures across the electromagnetic spectrum due to radiative processes such as synchrotron, inverse Compton (IC), bremsstrahlung, and Coulomb energy losses \cite{cola}. 

While there have been considerable efforts to study gamma-ray emission from dark matter annihilation in a variety of systems, e.g. \cite{Ackermann2010,Ackermann2015_dSph, charles,daylan, HESS_dSph,  HESS_GC}, a multiwavelength approach provides a complementary probe and in certain cases stronger constraints on dark matter properties \cite{storm, storm16}. The synchrotron emission from these particles is the result of ambient magnetic fields that accelerate the charged particles, causing them to emit radiation at radio wavelengths. The IC radiation peaks at X-ray frequencies and is the result of photons from various radiation sources such as the CMB and starlight being up-scattered by the relativistic particles. 

For a multiwavelength approach to indirect dark matter searches we focus on three main categories of astrophysical targets: galaxy clusters, local group dwarf galaxies, and other nearby galaxies (including the Milky Way galactic center). Galaxy clusters are the largest virialized objects in the universe and are highly dark matter dominated. These are enticing targets due to the large presence of dark matter as well as the presence of $\mu$G scale magnetic fields \cite{Feretti, BrunettiComaRadio, Giovanni}, enabling synchrotron processes. Dwarf spheroidal galaxies (dSphs) are also targets of great interest to dark matter searches. The proximity of the local group dwarfs along with their low luminosity and high concentration of dark matter make them prime targets for indirect dark matter searches \cite{mateo, Strigari_dSph}. Particularly, dwarf spheroidal galaxies generally lack high radio and X-ray emission, which allow us to place stronger constraints on dark matter properties by analyzing the synchrotron and IC radiation from dark matter annihilation. Other interesting targets for dark matter searches include galaxies such as M31 \cite{Dugger, USC_M31} or the Galactic center of the Milky Way \cite{RegisUllioGC,  laha, HESS_GC}. These systems are thought to be rich in dark matter, as well as to contain magnetic fields capable of producing synchrotron emissions from dark matter annihilation products. Particularly, reports of gamma-ray excesses in these systems \cite{AckermannM31, daylan} that could potentially be due to the presence of dark matter make these compelling targets, since a gamma-ray signal from dark matter should be accompanied by radio and X-ray signatures. A difficulty with these targets however, is the presence of other astrophysical processes that can create signatures similar to what we would expect to see from dark matter.

In order to model the multiwavelength DM signal, besides the relevant radiative processes there are additional important effects such as spatial diffusion of the charged particles that require greater study. In former studies of galaxy clusters for instance, the role of diffusion has been estimated to be negligible \cite{cola}, whereas in other systems such as dSphs it can not be ignored \cite{cola_draco}. The extent to which diffusion affects the analysis of a system is determined by factors including the physical size of the region, energy losses of the particles, and magnetic fields. For example, in larger environments such as galaxy clusters the particle byproducts of dark matter annihilation are able to lose all their energy within the region of study, whereas in smaller systems the energetic particles escape the system before fully radiating through synchrotron and IC processes. Additionally, the strong dependence on the magnetic field of synchrotron losses and diffusion effects means that uncertainties in the magnetic field must be examined before making assumptions on the role of diffusion.

To facilitate multiwavelength indirect dark matter searches in astrophysical systems, the main purpose of this paper is to introduce and describe the RX-DMFIT (Radio and X-ray - DMFIT) tool. RX-DMFIT is an extension of the DMFIT \cite{DMFIT} tool developed by Jeltema \& Profumo (2008) which is used for gamma-ray fitting. The RX-DMFIT code\footnote{https://github.com/alex-mcdaniel/RX-DMFIT} is publicly available and provides the user a tool with which to calculate the properties of secondary emission from dark matter annihilation due to synchrotron and IC processes. In particular, it relies on the DarkSUSY \cite{darksusy} Fortran package to provide the electron/positron injection spectrum for a given dark matter mass and annihilation channel. From the injection spectrum the RX-DMFIT tool calculates the emissivities and fluxes based on the user provided properties of the astrophysical system. Also, provided observational flux density data, RX-DMFIT can calculate dark matter particle constraints from synchrotron and IC radiation. The tool consists of 19 C++ files including 5 .h header files and interfaces with the DarkSUSY Fortran package. Integrations are carried out using the methods from the GNU Scientific Library \cite{GSL}. Users have the ability to specify a multitude of system parameters including physical size of the system, magnetic field strength, dark matter density profile, and diffusion properties among others. In all, RX-DMFIT has roughly 15 different physical parameters to be manipulated.

This paper is organized in the following manner. In section \ref{sec:radiation} we describe the analytic solution of the diffusion equation and subsequently derive the synchrotron and IC flux densities. In section \ref{sec:params} we assign and describe parameter values chosen for the models used in our analysis, which we then analyze using the RX-DMFIT tool in section \ref{sec:app} showing the effects of altering system components such as the role of diffusion and the magnetic field. In this section, we also demonstrate the use of the tool to place constraints on the DM particle cross-section using radio observations before presenting our conclusions in section \ref{sec:conclusion}. In this paper, we assume a $\Lambda$CDM universe with $H_0 = 70.4$ km s$^{-1}$ Mpc$^{-1}$, $\Omega_m = 0.272$, $\Omega_{\Lambda} = 0.73$. We note here that these cosmological parameters are fixed in RX-DMFIT, though they are readily accessible in the source code in case adjustments are desired.

\section{Radiation From DM Annihilation}\label{sec:radiation}

\subsection{Diffusion Equation}

In order to calculate the synchrotron and IC emission from DM annihilation, we must first obtain the equilibrium $e^{\pm}$ spectrum by solving the diffusion equation:
\begin{equation}\label{eq:diff}
\pd{}{t} \pd{n_e}{E}  = \grad  \left [D(E,\v{r}) \grad{ \pd{n_e}{E} } \right] +
 \pd{}{E}\left[b(E,\v{r}) \pd{n_e}{E} \right]
+Q(E,\v{r}).
\end{equation}
Here $\partial n_e/\partial E$ is the equilibrium electron density,  $Q(E, \v{r})$ is our electron source term, $D(E, \v{r})$ is the diffusion coefficient, and $b(E, \v{r})$ is the energy loss term. We assume equilibrium and seek a steady state solution, thus we set the time dependence on the left hand side of the equation to zero. Our source term is given by, 
\begin{equation}\label{eq:source}
Q\left(E, r\right) = \frac{\avg{\sigma v } \rho_{\chi}^2(r) }{2M_{\chi}^2 } \d{N}{E_{inj}},
\end{equation}
where we use the Fortran package DarkSUSY v5.1.2 to determine the electron/positron injection spectrum per dark matter annihilation event, $dN/dE_{inj}$, which is dependent on the DM particle mass, annihilation channel, and the source energy, $E$.

For the diffusion coefficient, we adopt a spatially independent form with a power law energy dependence. The RX-DMFIT tool includes two forms for the diffusion coefficient: a simplified power law in energy, and another that incorporates the degree of uniformity of the magnetic field. They are respectively:
\begin{subequations}
\begin{align}
D(E) &= D_0 E^{\gamma}  \label{eq:diffSimple}\\
D(E) &= D_0 \frac{d_B^{2/3}}{B_{avg}^{1/3}}E^{\gamma} \label{eq:diffAdj},
\end{align}
\end{subequations}
where $d_B$ is  the minimum uniformity scale of the magnetic field and $D_0$ is the diffusion constant \cite{cola, ColaBlasi, BlasiCola}.

In the full energy loss term we include contributions from synchrotron, inverse compton (IC), Coulomb, and bremsstrahlung losses. Each energy loss term is dependent on the energies of the electrons and positrons, as well as the magnetic field strength in the case of synchrotron losses and the CMB photon spectrum for IC losses. Additionally, the Couloumb and bremmstrahlung losses are dependent on the thermal electron density, $n_e$. The full energy loss expression is

\begin{equation}\label{eq:bloss}
\begin{split}
b(E,\v{r}) & = b_{IC}(E) + b_{Synch.}(E,\v{r}) + b_{Coul.}(E) + b_{Brem.}(E)\\
	& =  b_{IC}^0E^2 + b^0_{Synch.}B^2(r) E^2 \\
	&+ b_{Coul.}^0 n_e \left(1+\log\left(\frac{E/m_e}{n_e }\right)/75 \right)\\
	& + b^0_{Brem.}n_e  \left( \log\left(\frac{E/m_e}{n_e }\right) + 0.36 \right).
\end{split}
\end{equation}
Here $n_e$ is the mean number density of thermal electrons. For high energy $e^{\pm}$ the synchrotron and IC losses are dominant.

A general analytic solution for equation \ref{eq:diff} has previously been determined for the case of homogenous diffusion using the Green’s function method \cite{cola, ginzburg}, which in general can also be applied to non-stationary sources. We are interested in the steady-state solution, and following the notation of Colafrancesco et. al. (2006) \cite{cola} have a solution of the form,

\begin{equation}\label{eq:dndeeq}
\pd{n_e}{E}= \frac{1}{b(E,\v{r})}\int_E^{M_{\chi}}dE\p G\left(r,v(E)-v(E\p)\right)Q(E, \v{r}).
\end{equation}
where the Green's function,  $G\left(r,v(E)-v(E\p)\right)$,  is given by, 

\begin{multline}
G(r,  \Delta v ) = \frac{1}{\sqrt{4\pi \Delta v }}\sum_{n = -\infty}^{\infty} \left(-1\right)^n\int_0^{r_h}dr\p\frac{r\p}{r_n} \left(\frac{ \rho_{\chi}(r\p) }{\rho_{\chi}(r)}\right)^2\\ 
 \times\left[
\exp\left(-\frac{ (r\p - r_n)^2 }{4 \Delta v}	\right)	- 	\exp\left(-\frac{ (r\p + r_n)^2 }{4 \Delta v}	\right)	\right].
\end{multline}
As in previous work \cite{cola, BaltzEdsjo}, we impose the free escape boundary condition at the radius of the diffusion zone, $r_h$, using the image charge method with charges placed at  $ r_n = (-1)^nr + 2n r_h$.
Information about both the diffusion coefficient and energy loss terms have been incorporated into the $\Delta v = v(E)-v(E\p)$ term, where $v(E)$ is:
\begin{equation}
v(E) = \int_E^{M_{\chi}} d\tilde{E} \frac{D(\tilde{E})}{b(\tilde{E})}.
\end{equation}
Here $\sqrt{\Delta v}$ has units of length and gives the mean distance traveled by an electron as it loses energy between its source energy, $E\p$, and interaction energy, $E$. Note that in order to derive the Green's function for the diffusion equation using the method of Colafrancesco et. al. (2006) a spatially independent magnetic field is needed. For evaluation of the Green's function we approximate the energy loss term, $b(E, \v{r}) \approx b(E)$ by using an average magnetic field strength. That is, in equation \ref{eq:bloss} we take, 
\begin{equation}
b_{Synch.}(E) \approx  b^0_{Synch.}B^2_{avg} E^2.
\end{equation}
This approximation is used only in the evaluation of the Green's function, whereas for the energy loss term outside the integral of equation \ref{eq:dndeeq} we incorporate the full spatial profile of the magnetic field.

\subsection{Synchrotron}

The electrons and positrons produced as a result of dark matter annihilation produce multiwavelength emission through mulitple radiative processes. At radio frequencies, in the presence of reasonably strong magnetic fields (i.e. $B >B_{CMB} \simeq 3.25(1+z)^2 \mu G$) energy losses of the relativistic electrons and positrons are dominated by synchrotron radiation. From \cite{storm16, longair} we have the synchrotron power for a frequency $\nu$ averaged over all direction as:

\begin{equation}
P_{syn} \left(\nu, E , r\right) = \int_0^{\pi} d\theta \frac{\sin \theta}{2} 2\pi \sqrt{3}r_0 m_e c \nu_0 \sin\theta F\left(\frac{x}{\sin\theta}\right), 
\end{equation}
where $r_0 = e^2/(m_ec^2)$ is the classical electron radius, $\theta$ is the pitch angle, and $\nu_0 = eB/(2\pi m_e c)$ is the non-relativistic gyrofrequency. The $x$ and $F$ terms are defined as,

\begin{equation}
x \equiv \frac{2\nu \left(1+z\right)m_e^2}{3\nu_0 E^2},
\end{equation}

\begin{equation}
F(s) \equiv s\int_s^{\infty} d \zeta K_{5/3}\left( \zeta \right)1.25 s^{1/3}e^{-s}\left[648 + s^2\right]^{1/12},
\end{equation}
where $K_{5/3}$ is the Bessel function of order 5/3. The synchrotron emissivity at a frequency $\nu$ is found by folding the synchrotron power and electron equilibrium spectrum:

\begin{equation}
j_{syn} \left(\nu , r\right)= 2\int_{m_e}^{M_{\chi}} dE \d{n_{e}}{E}\left(E, r\right)P_{syn}\left(\nu, E, r \right).
\end{equation}
From this we calculate the integrated flux density spectrum, which we find by taking the line of 
sight integral of the emissivity to find the surface brightness, then subsequently integrate the surface brightness over the solid angle of the emission region. This gives us:

\begin{equation}
S_{syn} (\nu) = \int_{\Omega}d\Omega \int_{los} dl j_{syn} \left(\nu, l \right). 
\end{equation}
Approximating the target as a small region with much greater distance than size gives the final result:
\begin{equation}\label{eq:ssyn}
S_{syn} \approx \frac{1}{D_A^2} \int dr r^2 j_{syn}(\nu, r ),
\end{equation}
where $D_A$ is the angular diameter distance.

\subsection{Inverse Compton}

For regions with lower magnetic fields, the dominant radiative process is inverse Compton (IC) scattering of background photons, including most prominently the 2.73K Cosmic Microwave Background photons. Relativistic electrons and positrons from dark matter annihilation scatter the ambient CMB photons, producing a spectral peak between the soft to hard X-ray bands depending on the mass of the dark matter particle \cite{JP2012}. 
With the photon number density $n \left( \epsilon \right)$, and the IC scattering cross-section $\sigma\left( E_{\gamma} , \epsilon , E \right)$, the IC power is:

\begin{equation}
P_{IC} \left( E_{\gamma}, E \right) = c E_{\gamma}\int d\epsilon \: n \left( \epsilon \right) \sigma\left( E_{\gamma} , \epsilon , E \right).
\end{equation}

\noindent
Here $\epsilon$ is the energy of the target CMB photons, $E$ is the energy of the relativistic electrons and positrons, and $E_{\gamma}$ is the energy of the upscattered photons. $\sigma\left( E_{\gamma} , \epsilon , E \right)$ is given by the Klein-Nishina formula:
\begin{equation}
\sigma\left( E_{\gamma} , \epsilon , E \right) = \frac{3 \sigma_T}{4\epsilon \gamma^2} G\left( q, \Gamma \right),
\end{equation}

\noindent where $\sigma_T$ is the Thomson cross-section and $G(q, \Gamma)$ is given by \cite{Blumenthal}:

\begin{equation}
G (q, \Gamma) = \left[  2q\ln q + (1+2q)(1-q) + \frac{(2q)^2(1-q)}{2(1+\Gamma q)}  \right],
\end{equation}
where,
\begin{equation}
\Gamma = \frac{4\epsilon \gamma}{m_e c^2} = \frac{4\gamma^2 \epsilon}{E}, \qquad q = \frac{E_{\gamma}}{\Gamma(E-E_{\gamma})} 
\end{equation}
For this process, the range of values of $q$ is determined by the kinematics of the problem to be $1/\left( 4 \gamma^2 \right) \leq q \leq1$ \cite{cola, Blumenthal, Rybicki}.
As with the synchrotron emission, we find the local emissivity by folding the power with the electron equilibrium density, 
\begin{equation}
j_{IC} \left(E_{\gamma}, r\right)= 2\int_{m_e}^{M_{\chi}} dE \d{n_{e}}{E}\left(E, r\right)P_{IC}\left(E, E_{\gamma}\right),
\end{equation}
and the (approximate) integrated flux density is:
\begin{equation}
 S_{IC} \approx \frac{1}{D_A^2} \int dr r^2 j_{IC}(E_{\gamma}, r ),
\end{equation}


\section{Parameter Selection}\label{sec:params}
In the following sections we describe and assign the various parameters required to define our target, and present the results of radiation from DM annihilation as calculated by RX-DMFIT. We will demonstrate the use of RX-DMFIT by performing our analysis on three scales: A cluster scale model emulating the Coma cluster, where we assume a redshift  $z = 0.0232$ and diffusion zone $r_h = 415$ kpc  \cite{storm}; a dwarf spheroidal model similar to the Draco dwarf with redshift corresponding to a distance of 80 kpc \cite{dracoDist} and a diffusion zone $r_h = 2.5$ kpc \cite{cola_draco}; and finally a galactic scale model similar to M31 at a distance 780 kpc \cite{M31dist} and with a diffusion zone radius of $r_h = 30$ kpc borrowing from analysis of the Milky Way \cite{strong}.

\subsection{Magnetic Field Model}
The RX-DMFIT tool currently supports two magnetic field models. These are 

\begin{subequations}
\begin{align}
B(r) &= B_0 \: e^{-r/r_c} \label{eq:Bex} \\
B(r) &= B_0 \left[ 1+ \left(\frac{r}{r_{c}} \right)^2 \right]^{-1.5\beta\eta} \label{eq:Bbeta},
\end{align}
\end{subequations}
where $B_0$ is the central magnetic field strength and $r_c$ is the core radius of the target system. 

{\bf Clusters:}
The presence of large scale magnetic fields in galaxy clusters has been demonstrated through various methods such as observations of radio halos, purported inverse compton X-ray emission, and Faraday Rotation Measures (FRM) among others \cite{govoni}. The typical ranges that have been determined for magnetic field strength in non-cool-core clusters based on FRMs are $\sim$ 1-10 $\mu$G, whereas clusters with cool cores have been found to host magnetic fields in the range of $\sim$ 10-40 $\mu$G \cite{carilli}. In our analysis we explore both a ``non-cool-core'' (NCC) model and a ``cool-core'' (CC) model. A prototypical and well-studied NCC cluster is the Coma cluster, with a reported central magnetic field of $B_0 = 4.7 \mu $G and  $r_c = 291$ kpc \cite{bonafede}. For the CC cluster model, the Perseus cluster provides the prototypical example with a field strength $B_0 = 25 \mu $G \cite{Perseus} and core radius $r_c = 46$ kpc \cite{storm}. CC clusters typically have higher central fields with steeper profiles whereas the NCC clusters tend to host lower strength, shallow field profiles. These differences are generally attributed in part to major mergers of NCC clusters that destroy the cool core \cite{hudson, burns}. In both the NCC and CC systems we adopt the the beta model magnetic field profile of equation \ref{eq:Bbeta}. This choice of the profile is motivated by simulations \cite{murgia} along with observations of clusters such as Coma \cite{bonafede} that suggest magnetic fields in clusters scale with the thermal gas density which is often modeled with a beta-model \cite{cavaliere}. We also include the free-parameter $\eta$ as in previous cluster magnetic field modeling \cite{bonafede, vacca}. The $\beta$ and $r_c$ parameters are typically fit by X-ray observations \cite{Chen}, whereas $\eta$ is usually fit using FRMs \cite{vacca}. While the values for $\beta$ and $\eta$ are easily adjusted by the user in RX-DMFIT we will adopt $\beta = 0.75 $ and $\eta = 0.5$ throughout our calculations, noting that the effect of varying these parameters is minimal \cite{storm16}.

{\bf dSphs: }
Previous explorations of the magnetic field present in dSph galaxies show that any fields present would be relatively small, with most estimates for the magnetic field strength being $B_{\mu} \sim 1 \mu$G \cite{spekkens, cola_draco}, although some estimates are as large as $B_{\mu} \sim 2 \mu$G for dwarfs in the outer regions of the Milky Way magnetic field \cite{natarajanSegue}. For our purposes we will adopt the more conservative estimates of a central strength $B_{\mu} = 1 \mu G$. The spatial profile of magnetic fields in dwarfs is similarly poorly defined, leading us to adopt the simple exponential model of equation \ref{eq:Bex}. For the estimate of the core radius we take the half-light radius of Draco to be $r_c = 0.22$ kpc \cite{rstar}.
\newline
\indent {\bf Galaxies:}
The magnetic fields structure in galaxies is often considerably more complex than considered in this analysis. However, for our purposes we again employ the exponential model given by equation \ref{eq:Bex} for the magnetic field, while noting that a full treatment of the magnetic fields structure in galaxies can potentially impact the resulting synchrotron emission. Values for the magnetic field in the centermost region of M31 have been reported to be up to 15 $\mu$G \cite{hoernes}. Using a core radius of 10 kpc \cite{strong}, this value provides us with an average field strength of $\sim$ 4.8 $\mu$G in our model which is consistent with previous studies of M31 \cite {GandBeck}.

\subsection{Dark Matter Profile}
The DM profile modeling supports user-selection of the Navarro-Frenk-White (NFW) profile \cite{NFW96, NFW97}, as well as the Einasto profile \cite{Einasto, N04} in the forms,
\begin{align}
\text{NFW: } \: &\rho\left(r\right)  = \frac{\rho_{s} }{\left(\frac{r}{ r_{s} }\right) \left(1 + \frac{r}{r_{s} } \right)^2 }\\
\text{Einasto: } \: & \rho\left(r\right)  = \rho_s \exp \left\{-\frac{2}{\alpha} \left[ \left(\frac{r}{r_s}\right)^{\alpha} - 1 \right] \right\}  \label{eq:Einasto}.
\end{align}
In the RX-DMFIT code, users supply relevant characteristic density, $\rho_s$, and radius, $r_s$, as well as the $\alpha$ parameter for the Einasto profile. In this paper we will restrict our analysis to mainly make use of the NFW profile, and use the same NFW density and radius values for both the NCC and CC cluster models. The parameters chosen for each example system with references are summarized in table \ref{tab:i}.
\begin{table}[tbp]
\centering
\begin{tabular}{|l|c|c|c|}
\hline
  & $\rho_s$ (GeV/cm$^3$) &$r_s$ (kpc) & Ref.\\
\hline
Cluster & $ 0.0399$ &$404$& \cite{storm}\\
dSph& $1.40$ &$1$&\cite{cola_draco}\\
Galaxy& $ 0.184$ &$24.42$&  \cite{BuchCirelli}\\
\hline
\end{tabular}
\caption{\label{tab:i} Dark matter density parameters of each system for an NFW profile.}
\end{table}

\subsection{Diffusion Parameters}

Due to the lack of concrete values for diffusion in the different systems being studied here, we adopt the same initial parameters across our cluster, dwarf, and galaxy models. In the following sections we will vary these parameters and see to what extent the role of diffusion is important on different astrophysical scales.

For diffusion modeling in this paper we will restrict ourselves to the simple power law in equation \ref{eq:diffSimple}. Most values for appropriate $D_0$ are based on studies of the Milky Way and fall in the range of $10^{27}$ - $10^{29}$ cm$^2$s$^{-1}$ \cite{webber, BaltzEdsjo, BuchCirelli}. Constraints on the Milky Way diffusion parameters have been determined based on measured B/C data in the galaxy \cite{webber, maurin}. We can also consider the $D_0$ parameter in terms of its relation to the inhomogeneity of the magnetic field in order to understand how it scales with the size of the system. Estimates for the diffusion constant can be found by assuming $D_0 \sim V_L L$, where $V_L$ is the amplitude of the turbulent velocity and $L$ is the scale of the turbulent motions \cite{lazarian, TeslaXray}. Scaling these parameters for dwarf spheroidals, normal galaxies, and galaxy clusters provides diffusion constant values compatible with the range above. Furthermore, the overall size of the system and the magnetic field strength play a role in whether or not diffusion has a significant impact on the resulting emission. In cluster sized systems, the length scale over which the electron/positrons lose their energy, given by $\sqrt{\Delta v}$, will typically be less than the diffusion zone $r_h$. In contrast, relativistic particles in smaller systems such as Milky Way sized galaxies and dwarf spheroidals will be able to escape the diffusion zone before radiating their energy. In each of these systems, greater magnetic field strength will result in the relativistic particles radiating their energy more quickly before escaping the system. These effects can also be considered in terms of the relevant timescales for each energy loss process in comparison to the timescale for diffusion, with a useful example provided in figure A.3 of Appendix A of \cite{cola}. While there are a lack of studies into values for the diffusion constant in other astrophysical systems, the range of $10^{27}$ - $10^{29}$ cm$^2$s$^{-1}$ provides reasonable estimates that we can apply to our models. 

Following previous work \cite{cola} we assign $\gamma = 1/3$ and take the parameter values for the energy loss coefficients in equation \ref{eq:bloss} to be $b_{syn}^0 \simeq 0.0254$, $b_{IC}^0 \simeq 0.25$, $b_{brem}^0 \simeq 1.51$, and $b_{Coul}^0 \simeq 6.13$, all in units of $10^{-16}$ GeV/s. Additionally, we also must select appropriate values for the average thermal electron density, $n_e$. For our cluster models we take $n_e \approx 10^{-3}$ \cite{storm}, $n_e \approx 10^{-6}$ \cite{cola_draco} for dwarf spheroidals, and estimate $n_e \approx 0.1$ \cite{ferriere} for our galaxy model.



\section{Application and Results}\label{sec:app}
\subsection{Diffusion Effects}

\begin{figure}[tbp]
\centering 
\includegraphics[width=\textwidth]{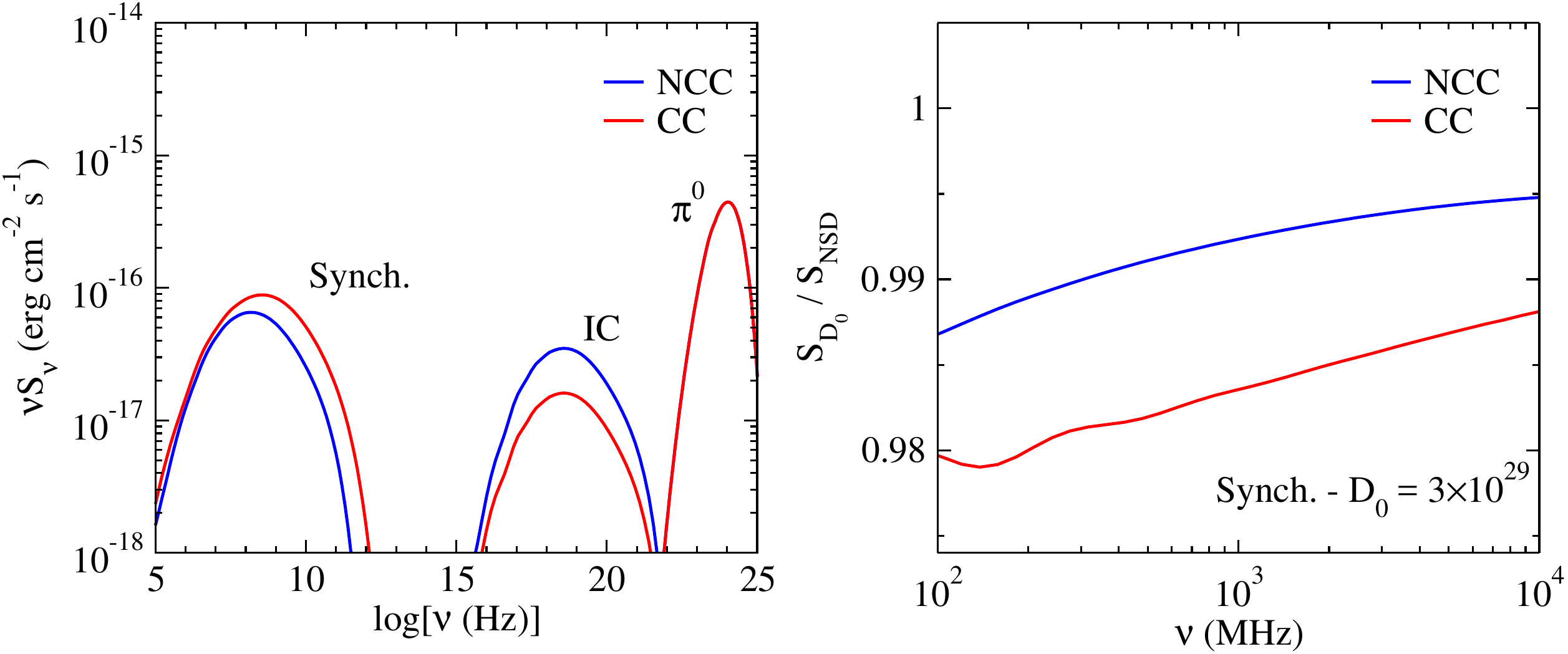}
\caption{\label{fig:sedDCCvNCC}Left: SED comparing the ``non-cool-core'' (NCC) and ``cool-core'' (CC) cluster models assuming a $b\bar{b}$ final state with $M_{\chi} = 100$
GeV. Here we use only the limit of no spatial diffusion (NSD). Right: Ratio of synchrotron flux with  $D_0 = 3\times 10^{28}$ cm$^{-3}$s$^{-1}$ over the NSD limit for the NCC and CC cluster models.}
\end{figure}

\begin{figure}[tbp]
\centering 
\includegraphics[width=0.6\textwidth]{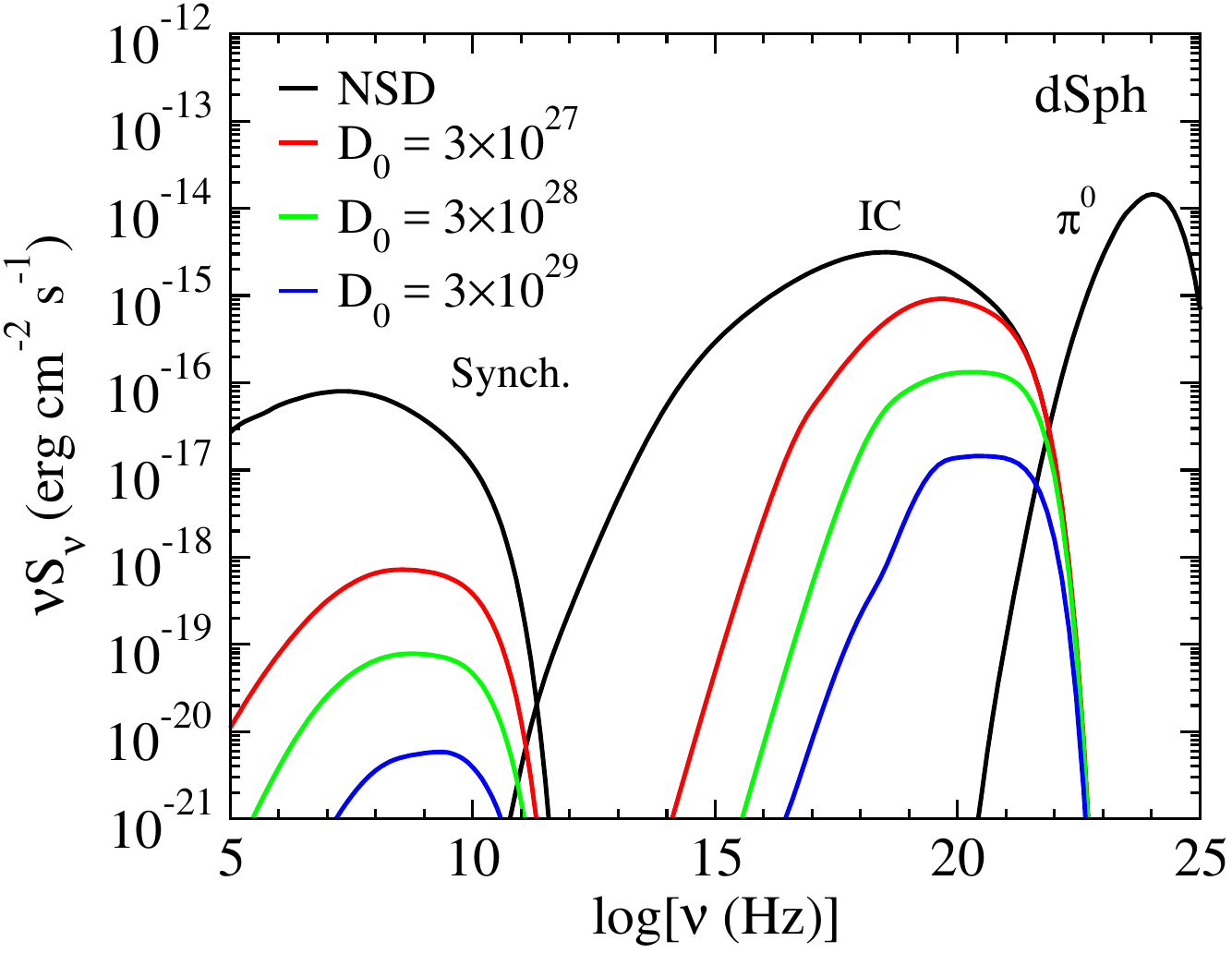}
\hfill
\caption{\label{fig:dSED_Diff} SED of the dwarf model assuming a $b\bar{b}$ channel with $M_{\chi} = 100$ GeV for multiple values of $D_0$ in cm$^2$ s$^{-1}$. }
\end{figure}

\begin{figure}[tbp]
\centering 
\includegraphics[width=0.6\textwidth]{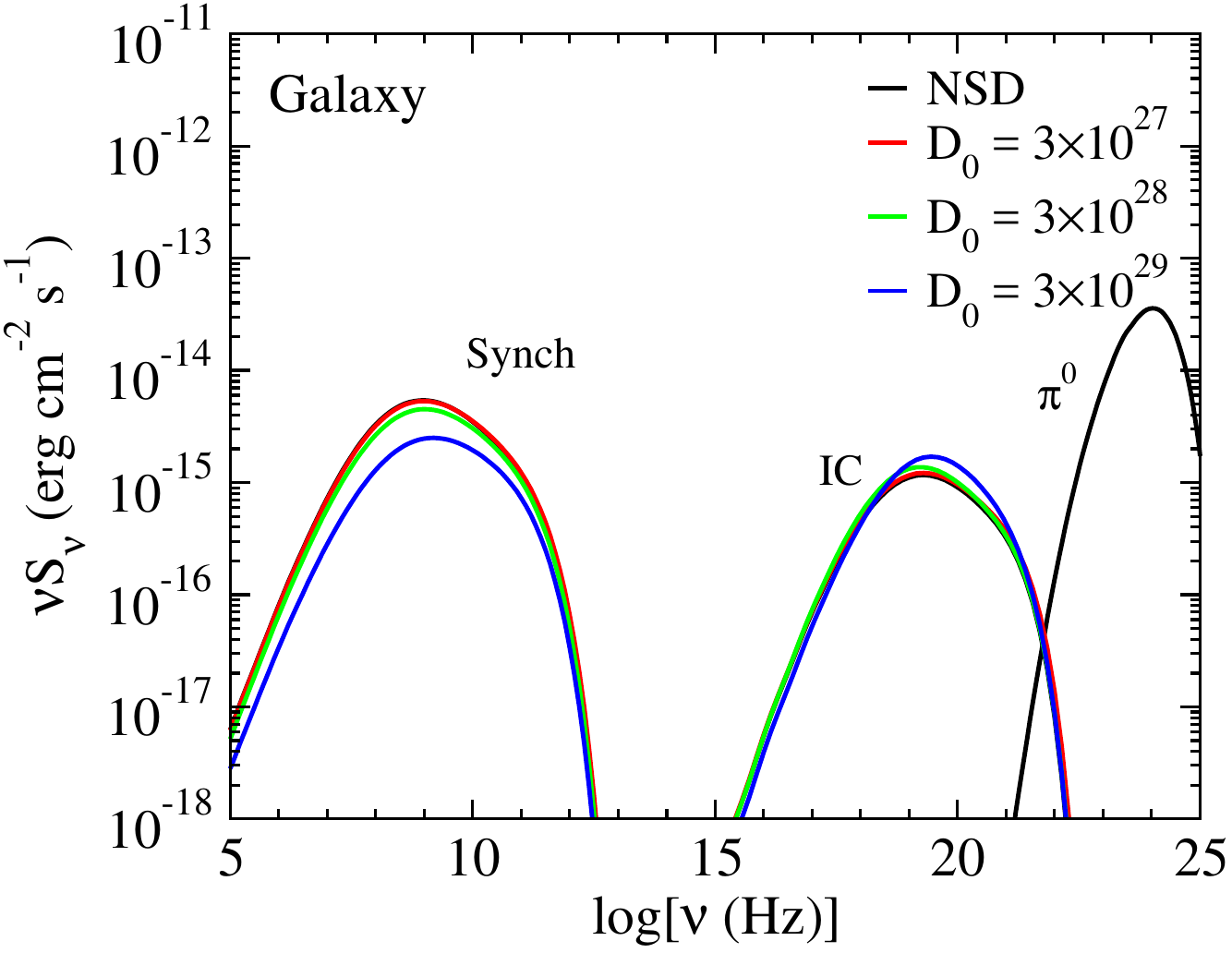}
\hfill
\caption{\label{fig:gSED_Diff} SED of the galaxy model assuming a $b\bar{b}$ channel with $M_{\chi} = 100$ GeV for multiple values of $D_0$ in cm$^2$ s$^{-1}$. Note that in the plot there is almost no noticeable difference between the NSD limit and the lowest diffusion values of $D_0 = 3 \times 10^{27} -3\times 10^{28}$ cm$^{2}$s$^{-1}$.}
\end{figure}

\begin{figure}[tbp]
\centering 
\includegraphics[width=0.6\textwidth]{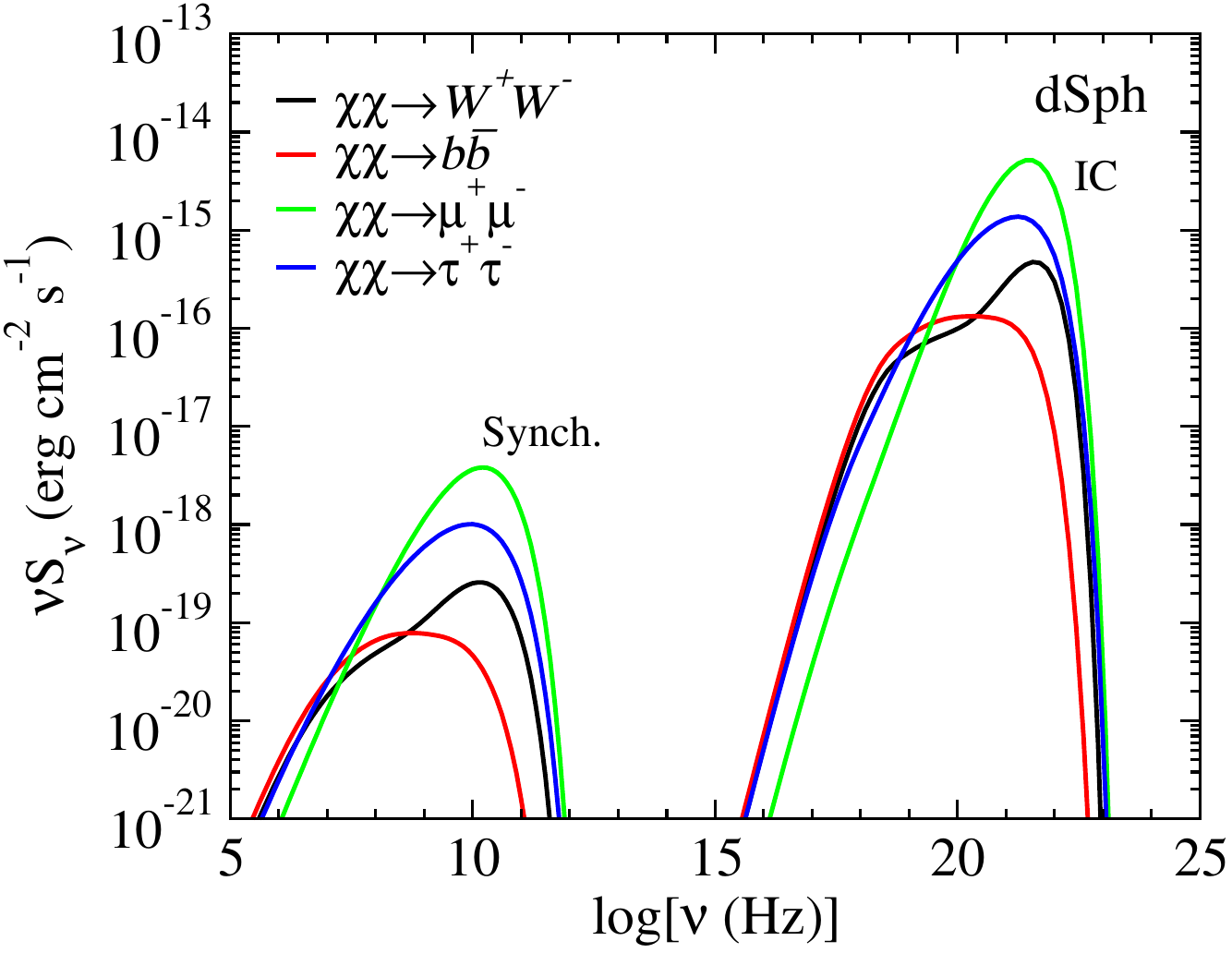}
\caption{\label{fig:SED_dch}SED of the dwarf model with $M_{\chi} = 100$ GeV for various annihilation channels assuming a diffusion constant of $D_0 = 3\times 10^{28}$ cm$^{-3}$s$^{-1}$. }
\end{figure}

\begin{figure}[tbp]
\centering 
\includegraphics[width=0.6\textwidth]{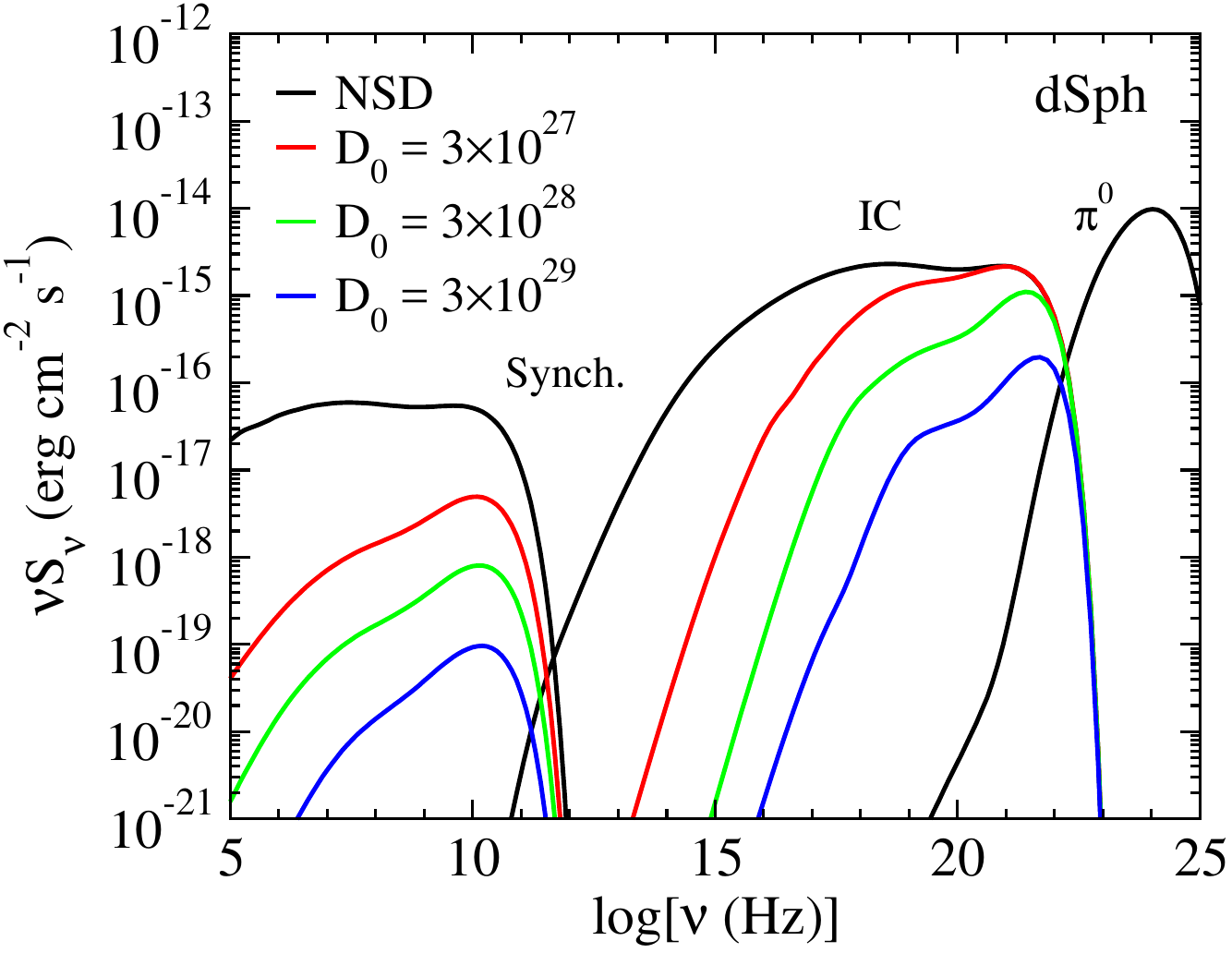}
\caption{\label{fig:SED_WW} SED of the dwarf model with $M_{\chi} = 100$ GeV and DM annihilation through the $W^+W^-$ channel for multiple values of $D_0$ in cm$^2$ s$^{-1}$. }
\end{figure}

We show the results of the SED and emissivity calculations using the RX-DMFIT tool. In figures \ref{fig:sedDCCvNCC}, \ref{fig:dSED_Diff}, and \ref{fig:gSED_Diff} we show the multiwavelength SED for each of of our main systems, assuming the $b\bar{b}$ annihilation channel dominates and including contributions from IC and synchrotron processes with various values for the diffusion constant $D_0$. To compare with the expected synchrotron and IC fluxes, in figures \ref{fig:sedDCCvNCC}, \ref{fig:dSED_Diff}, \ref{fig:gSED_Diff}, and \ref{fig:SED_WW} we also include the expected prompt gamma-ray emission due to the decay of neutral pions. Note that the gamma-ray emission is not affected by the magnetic field or diffusion parameters, simplifying the gamma-ray flux calculation (see for instance \cite{cola, RegisUllioGC}). For clarity, we do not include the gamma-ray fluxes in the SEDs of figures \ref{fig:SED_dch} and \ref{fig:SEDBPanel}.

Figure \ref{fig:sedDCCvNCC} shows a comparison of the SED for our CC and NCC cluster models. The CC model contributes more from synchrotron radiation due to its stronger magnetic field, whereas the decreased synchrotron emission in the NCC model leads to comparatively higher IC emissions. In both the CC and NCC models we do not observe significant impact of spatial diffusion for even the largest diffusion values of $D_0 = 3 \times 10^{29}$ cm$^{2}$s$^{-1}$ , which is consistent with previous estimations of the diffusion effect in galaxy clusters \cite{cola}. To help illustrate this point, in the right panel of figure \ref{fig:sedDCCvNCC} we show the ratio of flux density from synchrotron radiation in our cluster models with diffusion versus without diffusion over a range of frequencies. In both the CC and NCC models there is at most a $\sim$ 2\% decrease when considering our highest diffusion strength. In the case of dSphs, we see in figure \ref{fig:dSED_Diff} that diffusion at all included $D_0$ values plays a significant role in decreasing the total emission of both the synchrotron and IC radiation as the relativistic particles escape the diffusion region before radiating. In figure \ref{fig:gSED_Diff} we show the SED of our galaxy model. Here we observe a decrease in synchrotron emission at each $D_0$ value, however this is considerably less than in the dwarf model. For instance, the lowest diffusion constant value $D_0 = 3 \times 10^{27}$ cm$^{2}$s$^{-1}$ yields an essentially negligible decrease in synchrotron emission. Even at the highest value of $D_0 = 3 \times 10^{29}$ cm$^{2}$s$^{-1}$ there is only about a factor of two decrease in the synchrotron emission, in contrast to the roughly three order of magnitude decrease in the dwarf model for this diffusion value. We also note that the decrease in synchrotron emission is accompanied by a slight increase in the IC emission for our galaxy model. As the relativistic particles diffuse into regions of diminished magnetic field within the diffusion zone, IC emission scattering from the uniform CMB photon distribution becomes the dominant form of radiation.

We also consider a variety of particle models for dark matter annihilation wherein different channels dominate. In figure \ref{fig:SED_dch} we show the SED for our dwarf system under various assumptions for the DM annihilation channel. We note a harder spectrum for the leptonic $\mu^+\mu^-$ and $\tau^+\tau^-$ states than for the $b\bar{b}$ state, and a flatter spectrum for the $W^+W^-$ state. While the leptonic states have spectra that tend to slant more towards higher energies than the $b\bar{b}$ channel, the $W^+W^-$ channel combines aspects of both the leptonic spectra and the $b\bar{b}$ spectra due to the $W^+W^-$ decay into pions and leptons, resulting in a flattened spectral profile. Furthermore, as seen in figure \ref{fig:SED_WW}, increased diffusion tends to diminish this effect as the hard spectrum of the $W^+W^-$ channel becomes more prominent. The predicted SED is also affected by other properties of the dark matter particle model such as the cross-section and particle mass. Changing the DM particle cross-section only changes the overall normalization since the emission is directly proportional to the $\avg{\sigma v}$ by equation \ref{eq:source}. Varying the DM particle mass on the other hand will affect the shape and location of the spectrum, with higher $M_{\chi}$ values producing harder spectra.

\begin{figure}[tbp]
\centering 
\includegraphics[width=\textwidth]{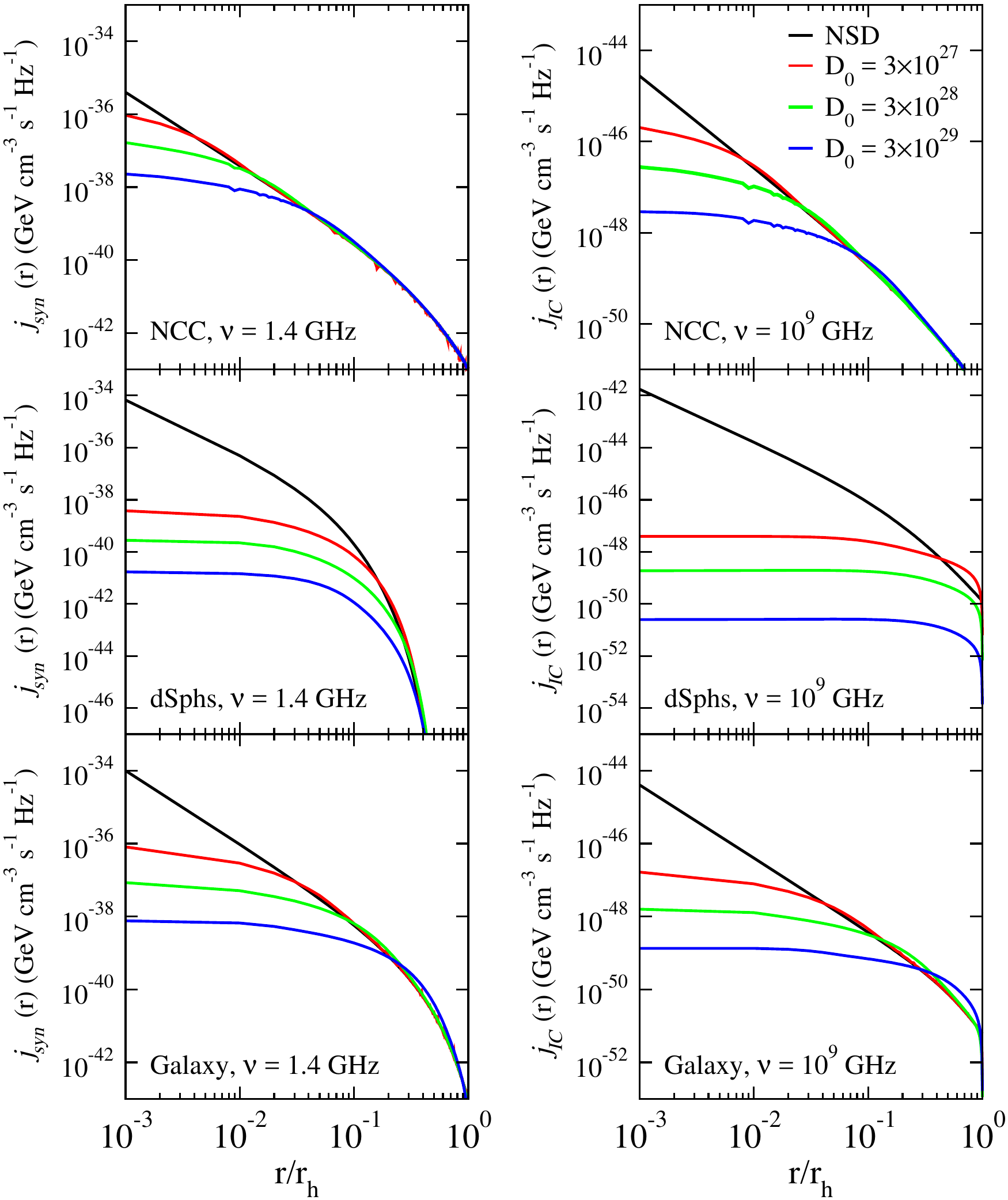}
\caption{\label{fig:EmissD} Predicted synchrotron ({\it left}) and IC ({\it right}) local emissivity profiles for the NCC, dSph, and galaxy models for an observing frequency $\nu$ (note that $10^9$ GHz $\approx 4$ keV) and various $D_0$ values. Our particle model assumes a dominant $b\bar{b}$ final state and a mass $M_{\chi} = 100$ GeV.}
\end{figure}

Diffusion effects can be seen more clearly by looking at the spatial local emissivity profile for synchrotron and IC emission. In figure \ref{fig:EmissD} we show the synchrotron and IC emissivity profiles for our NCC, dwarf, and galaxy models with various diffusion constant values. In our NCC model, introducing diffusion causes a slight decrease in the innermost region of the cluster which quickly returns to the NSD limit. For instance in the case of the the highest diffusion value of $D_0 = 3 \times 10^{29}$ cm$^{2}$s$^{-1}$
 the synchrotron profile approaches the NSD limit at $\sim$ 10 kpc and the IC emission reaches the NSD limit at $\sim$ 40 kpc. Furthermore, in neither case do we observe a considerable increase in emission along the profile. The NCC emissivity profiles are consistent with the lack of variation observed in the SEDs for the different $D_0$ values.

For our dwarf and galaxy models, including diffusion results in a large decrease in both synchrotron and IC emission for the central regions of each system. This depletion of emission is greater in the dwarf model than in the galaxy model, consistent with the SEDs of each system. We also note that diffusion leads to a slight excess in synchrotron emission in the outer regions of our dwarf system for the lower $D_0$ values. This excess is also present in the galaxy model for every $D_0$ value shown and for a larger portion of the diffusion zone. For instance, with a diffusion constant value of $D_0 = 3 \times 10^{27}$ cm$^{2}$s$^{-1}$ the synchrotron emission of the dwarf reaches the NSD limit at $\sim$ 0.5 kpc in comparison to $r_h$ = 2.5 kpc, whereas the the galaxy model reaches NSD limit at $\sim$ 0.9 kpc compared to $r_h = 30$ kpc. Both models also exhibit a flattened IC emission profile. In contrast to the synchrotron emission that depends on the radially dependent magnetic field, the IC emission depends on the spatially constant CMB photon distribution, leading to a flatter emission profile as the relativistic particles diffuse outward. While the dwarf model yields a slight excess of IC emission for the lowest diffusion strength, the galaxy model has a small excess in the outer regions for all diffusion values, providing the increase in IC emission observed in figure \ref{fig:gSED_Diff}.


\subsection{Magnetic Fields}
\begin{figure}[tbp]
\centering 
\includegraphics[width=\textwidth]{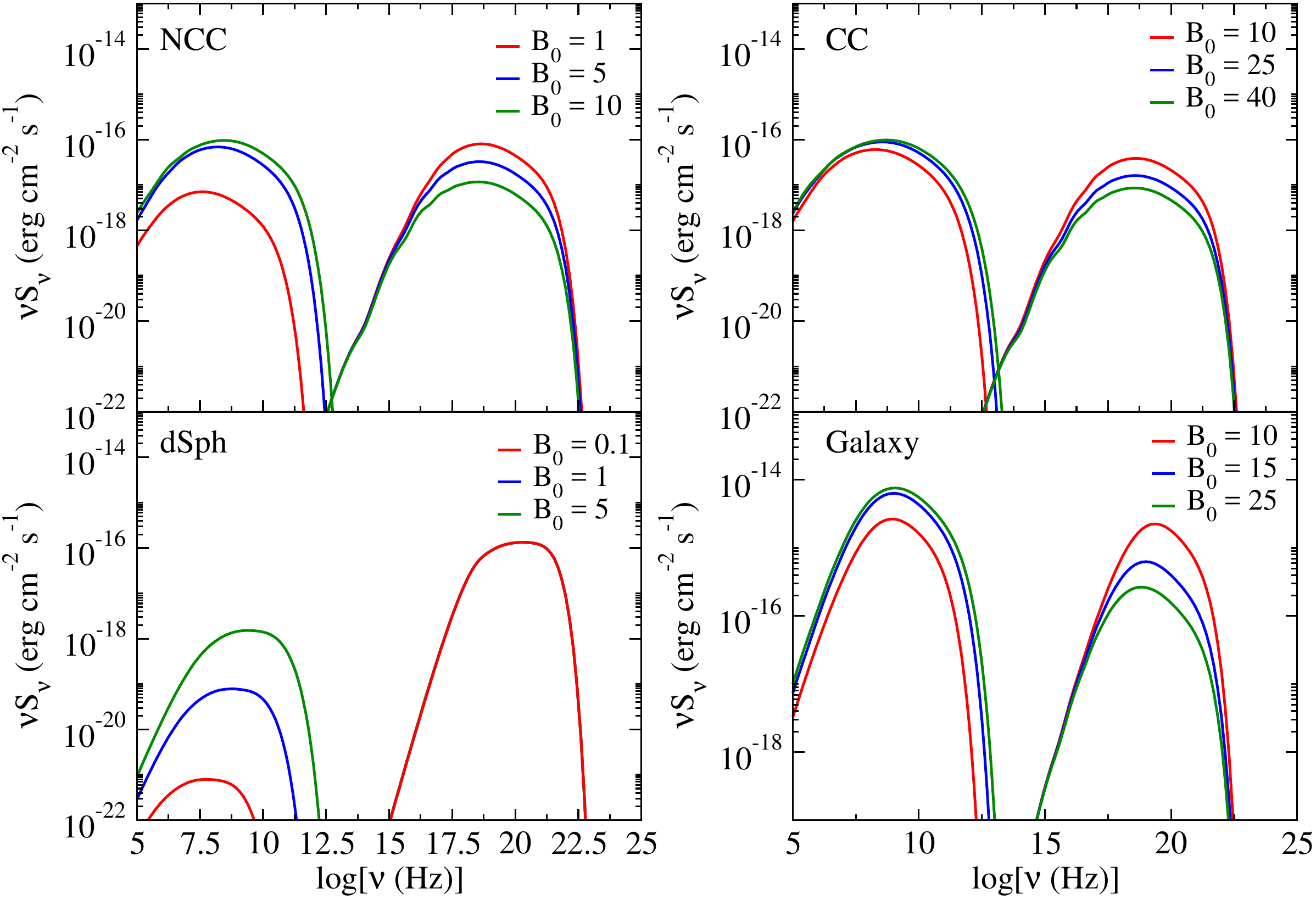}
\caption{\label{fig:SEDBPanel} SED for each model assuming a $b\bar{b}$ final state, with $M_{\chi} = 100$ GeV, and $D_0 = 3\times 10^{28}$ cm$^{2}$s$^{-1}$ (except for the NCC and CC models, in which diffusion is ignored) for multiple values of $B_0$ in $\mu$G. In the case of the dSph. model, while each $B_0$ value is shown there is no discernable difference in the IC SED.}
\end{figure}

Our ability to detect radio signals of from dark matter annihilation depends significantly on the magnetic field present in the system. In figure \ref{fig:SEDBPanel} we again show the multiwavelength SED for each of our models, this time varying central magnetic field strength in each case. We assume a diffusion constant value of  $D_0 = 3\times 10^{28}$ cm$^{2}$s$^{-1}$ for the dwarf and galaxy models, and assume no spatial diffusion for the NCC and CC cluster models. 

In each model, the magnetic field strength drastically impacts the the total synchrotron emission. For instance, decreasing the field strength on the dwarf model from $B_0 = 1$  $\mu$G to $B_0 = 0.1$ $\mu$G causes a decrease in the synchrotron radiation by roughly two orders of magnitude. For IC emissions, all of our models except the dSphs show significant dependence on the magnetic field strength, although with an inverse relationship. That is, lower magnetic field strengths in the galaxy and cluster systems lead to IC processes making up a greater portion of the total energy loss of the electrons and positrons. So while IC losses do not explicitly depend on magnetic field strength, systems with lower magnetic fields provide greater potential for IC radiation. For the NCC cluster model, we see that an order of magnitude increase in the magnetic field from  $B_0 = 1$ to  $B_0 = 10$ roughly translates into an even greater increase in radio emission, while decreasing the IC emission. In the CC cluster model there is less of a dependence on the central field strength, as shown by only a factor of $\sim$ 2 increase in synchrotron emission and factor of $\sim$ 4 decrease in IC emission from a factor of 4 increase in the magnetic field strength from $B_0 = 10$  $\mu$G to $B_0 = 40$  $\mu$G. The weaker dependence on the central magnetic field in the CC clusters versus the NCC cluster can be attributed to the smaller core radius of CC clusters. The steeper profiles of the CC clusters lead to a greater share of the synchrotron emission being confined to the inner regions of the clusters in comparison to the NCC clusters, meaning that altering the central field strength will have a lesser impact on the total emission in CC clusters than in NCC clusters.

\subsection{Dark Matter Constraints from Synchrotron Radiation}
Limits on the DM cross-section can also be determined using observed diffuse radio emission. To do this, we note that the flux density from dark matter given by equation \ref{eq:ssyn} is directly proportional to the thermal averaged DM particle cross-section through the source term given in equation \ref{eq:source}. Thus we can express the flux density as: 
\begin{equation}\label{eq:flux_constraint}
S_{\chi} = \frac{\avg{\sigma v}}{M_{\chi}^2}S_{\chi}\p, 
\end{equation}
where we have simply extracted the $\avg{\sigma v}$ dependence from the calculated flux density due to DM annihilation. We can then compare this quantity to an observed flux density for the system we are modeling and derive a constraint on the dark matter particle cross-section from, 
\begin{equation} \label{eq:crosssec}
\avg{\sigma v} = \frac{S_{obs}}{S_{\chi}\p}M_{\chi}^2.
\end{equation}
\begin{table}[tbp]
\centering
\begin{tabular}{|c|c|c|c|c|c|c|c|c|}
\hline
 $r_h$ (kpc)  & $d$ (kpc)    &   $D_0$ (cm$^2$s$^{-1})$    &    $\gamma$     &      $B_0$ ($\mu$G)    &      $\eta$    &    $\rho_s$ (GeV/cm$^3$)   &    $r_s$ (kpc)    &  $\alpha$\\
\hline
 $1.6$ & 23 &$3\times 10^{26}$& $0.7$ &$2$& $0$ & $6.6$ &$0.15$  & $1/3$\\
\hline
\end{tabular}
\caption{\label{tab:segue} Parameter selection for Segue I model.}
\end{table}

Here we present a practical example using RX-DMFIT wherein we derive dark matter constraints using radio data reported in Natarajan et. al. (2015) \cite{natarajanSegue} from $\nu = 1.4$ GHz observations of the Segue I dwarf galaxy with the Green Bank Telescope. From their analysis they find an upper limit flux density of $\sim 0.57$ Jy for a region of radius $\sim 4^{\circ}$. The physical parameters that we input into RX-DMFIT are taken from Natarajan et. al. (2015) \cite{natarajanSegue} and are summarized table \ref{tab:segue}, with any parameters that are not listed unchanged from our earlier dwarf model. Note that for consistency with Natarajan et. al. (2015) \cite{natarajanSegue}, we set $\beta$ (or equivalently, $\eta$) equal to zero in order to establish a constant magnetic field and employ the Einasto profile of \ref{eq:Einasto}, and thus include the $\alpha$ parameter. In addition to the fairly low diffusion value of $D_0 = 3\times 10^{26}$ cm$^{2}$s$^{-1}$, we also consider a greater diffsuion constant value of $D_0 = 3\times 10^{28}$ cm$^{2}$s$^{-1}$.

\begin{figure}[tbp]
\centering 
\includegraphics[width=0.6\textwidth]{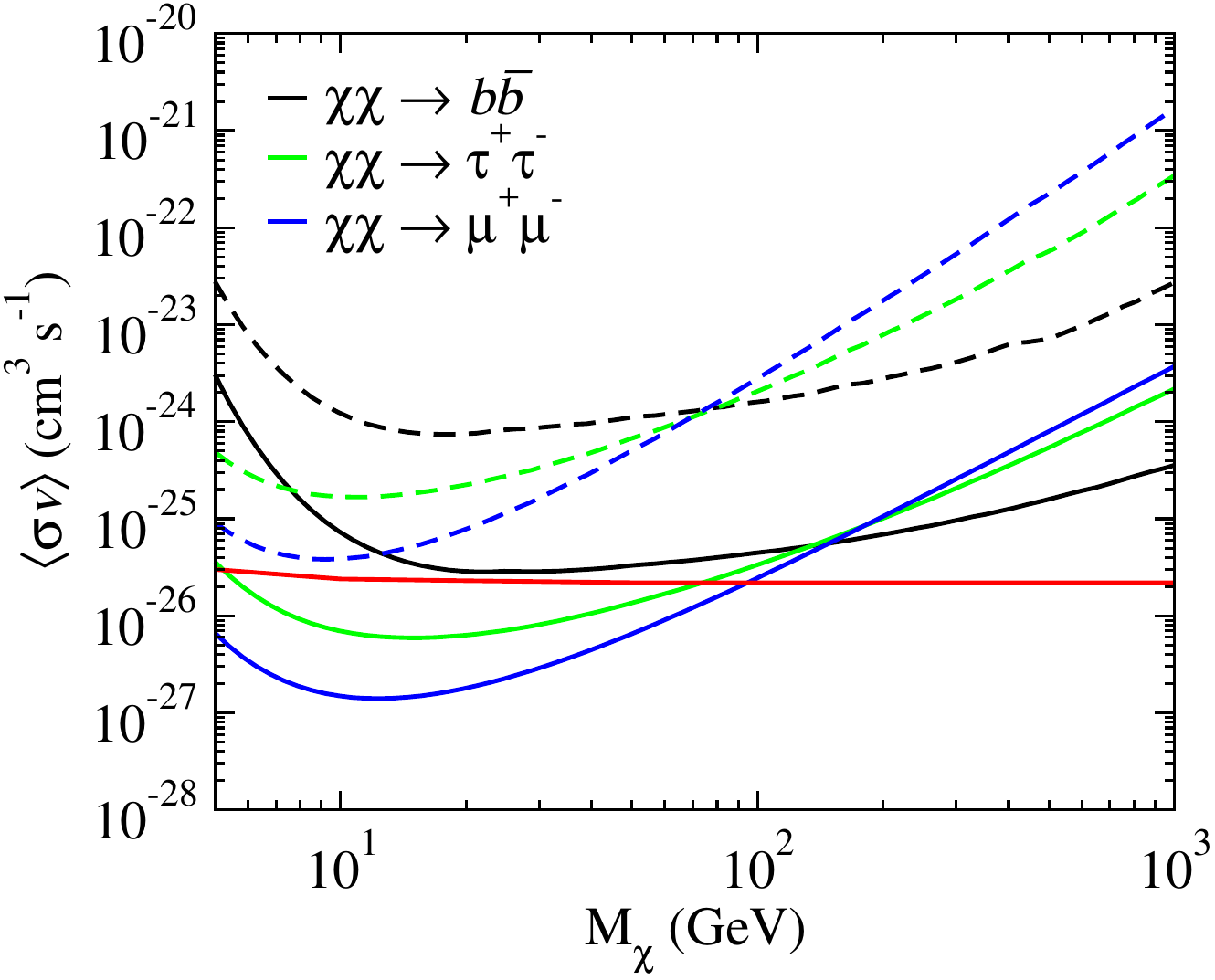}
\hfill
\caption{\label{fig:ExCurve_dD} Constraints on the DM particle cross-section from radio observations of the Segue I dwarf
galaxy for various annihilation channels and diffusion strengths. The solid lines indicate a diffusion
constant of  $D_0 = 3\times 10^{26}$ cm$^{2}$s$^{-1}$, whereas the dashed lines indicate $D_0 = 3\times 10^{28}$ cm$^{2}$s$^{-1}$. The red line is the thermal relic cross-section from Steigman et. al. (2012) \cite{Steigman}. }
\end{figure}

\begin{figure}[tbp]
\centering 
\includegraphics[width=\textwidth]{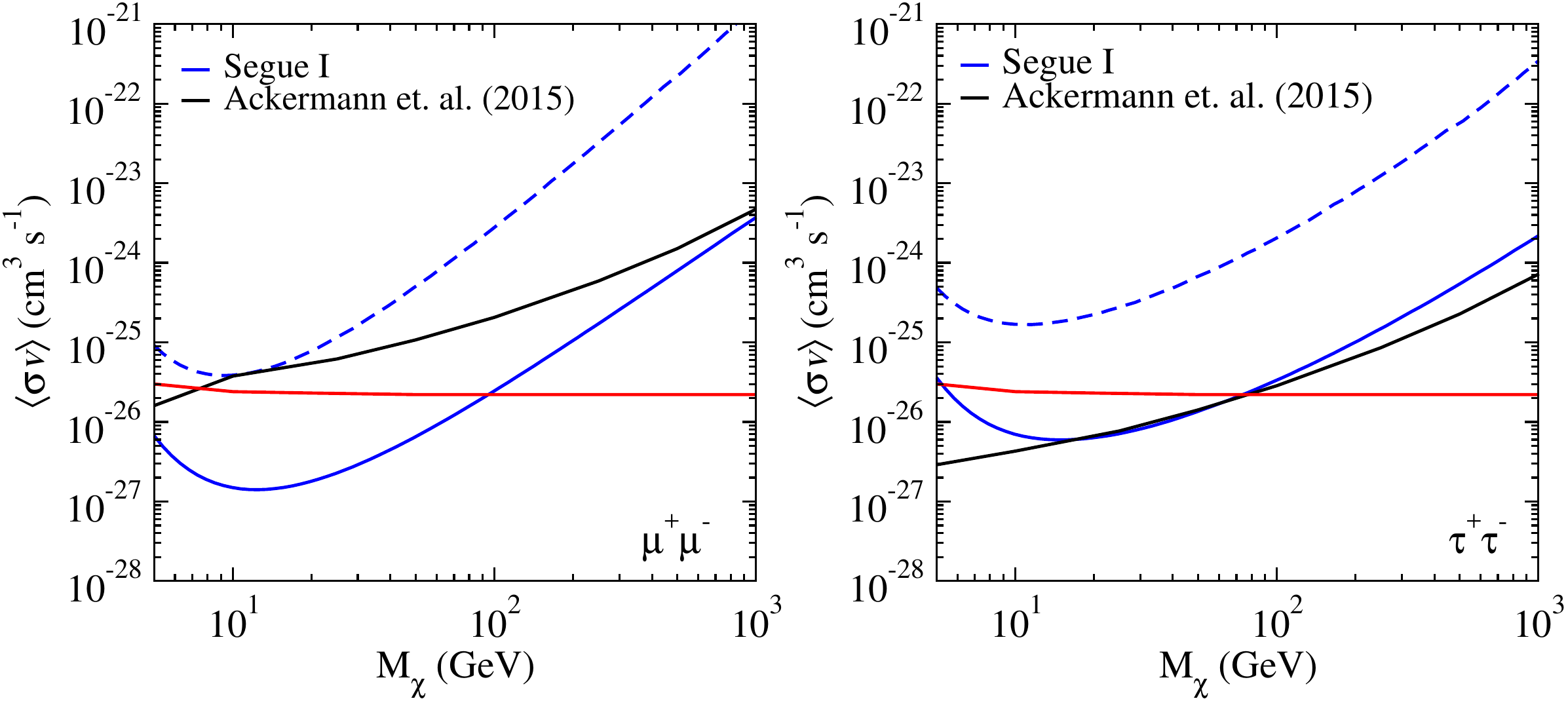}
\hfill
\caption{\label{fig:ExCurve_Fermi} Constraints on the DM particle cross-section from radio observations of the Segue I
dwarf galaxy in comparison to limits from Fermi gamma-ray data. The solid blue lines indicate a
diffusion constant of $D_0 = 3\times 10^{26}$ cm$^{2}$s$^{-1}$, whereas the dashed blue lines indicate  $D_0 = 3\times 10^{28}$ cm$^{2}$s$^{-1}$. The red line is the thermal relic cross-section from Steigman et. al. (2012) \cite{Steigman}. }
\end{figure}

Figure \ref{fig:ExCurve_dD} shows the upper limits on the annihilation cross-section for a variety of annihilation channels with and without diffusion. As we saw in the SED plots (see figures \ref{fig:dSED_Diff} and \ref{fig:SED_WW}), diffusion has a significant impact on the expected radio synchrotron emission in dwarf spheroidal galaxies, and in turn, a significant impact on the strength of the constraints  that can be placed on the DM particle. Increasing the diffusion constant from $D_0 = 3\times 10^{26}$ cm$^{2}$s$^{-1}$ to $D_0 = 3\times 10^{28}$ cm$^{2}$s$^{-1}$ weakens the constraints by an order of magnitude, and thus should not be neglected for our dwarf system. We find the strongest constraints for annihilation through the $\mu^+\mu^-$ and $\tau^+\tau^-$ channels, both of which reach below the thermal relic cross-section value for WIMP masses $M_{\chi} \leq 100$ GeV under weak diffusion
assumptions. 

These constraints are competitive with previous studies of dark matter in dSphs using
Fermi gamma-ray data \cite{Ackermann2015_dSph} which provides some of the strongest dark matter constraints from gamma-rays to date. In figure \ref{fig:ExCurve_Fermi} we compare the constraints placed on the dark matter cross-section from the combined gamma-ray observations of 25 Milky Way dSphs with six years of Fermi data \cite{Ackermann2015_dSph}. For $\tau^+\tau^-$ final states, weak diffusion, and masses around 10 GeV, the constraints are very similar. In the case of $\mu^+\mu^-$ dominated final states, the radio approach provides similar constraints for masses near 10 GeV in the case of high diffusion where $D_0 = 3\times 10^{28}$ cm$^{2}$s$^{-1}$ . With our lower value of $D_0 = 3\times 10^{26}$ cm$^{2}$s$^{-1}$
radio constraints are stronger for masses 5 GeV $\leq M_{\chi} \leq $ 1000 GeV including improvement upon the gamma-ray constraints by greater than an order of magnitude for masses 5 GeV $\leq M_{\chi} \leq$100 GeV. From these constraints we determine that dSphs are viable targets for indirect searches from dark matter annihilation by way of radio observations, and note that our results here are compatible with other radio constraints on dark matter annihilation. For instance, in the case of the Draco dwarf limits on the dark matter cross-section are in the range of $\avg{\sigma v} \sim 10^{-25}$ cm$^{2}$s$^{-1}$ \cite{spekkens, cola_draco}. Other radio constraints from the analysis of several dSphs \cite{Regis_dSphs} also are similar to the constraints found in this paper.

We are also interested in deriving limits on the dark matter WIMP models using X-ray observations. In the case of galaxy clusters, future hard X-ray observations have the potential to contribute significantly to dark matter constraints \cite{JP2012}. Additionally, Jeltema \& Profumo (2008) \cite{TeslaXray} have demonstrated that current and future X-ray observations of dwarf spheroidals can provide limits comparable and potentially better than limits from gamma-rays in a similar mass range as that for radio observation. However, these results rely on favorable assumptions for diffusion. More recently, there is deep X-ray data of the Draco dwarf that has been used for constraining dark matter decay \cite{Draco3.5} that can potentially provide stronger constraints on dark matter annihilation than those in \cite{TeslaXray} while making fewer assumptions about the diffusion and energy loss processes. In order to better understand the feasibility of obtaining dark matter constraints from X-rays in dwarfs we must take the X-ray background into account. For instance, recent Chandra results report cosmic X-ray background fluxes of $4.55^{+0.03}_{-0.03} \times 10^{-12}$ erg cm$^2$ s$^{-1}$ deg$^{-2}$ for the 1-2 keV ($\sim2.4-4.8 \times 10^{17}$ Hz) energy range and $2.034^{+0.005}_{-0.006} \times 10^{-11}$ erg cm$^2$ s$^{-1}$ deg$^{-2}$ for the 2-10 keV ($\sim 4.8-24.0 \times 10^{17}$ Hz) range \cite{Chandra}. From figure \ref{fig:SED_dch} we see that the predicted X-ray fluxes from DM annihilation in these energy ranges and for a 100 GeV DM particle are on the order of $\sim 10^{-16} - 10^{-14}$ erg cm$^2$ s$^{-1}$, depending on annihilation channel. The $\sim 2-5$ order of magnitude excess of the X-ray background over the predicted DM flux suggest that only conservative constraints can be placed without  an improved understanding of the X-ray background or deeper X-ray observations.

\section{Conclusion}\label{sec:conclusion}
We have presented RX-DMFIT, a new tool to analyze synchrotron and IC emission due to DM annihilation for the purposes of astrophysical indirect detection of dark matter. We considered four model systems: a ``non-cool-core'' as well as a ``cool-core'' galaxy cluster, a dwarf model, and a galaxy model. We discussed in detail the relevant astrophysical processes, namely diffusion of the charged particle byproducts of  DM annihilation, magnetic field modeling, and radiative energy loss processes.  We then used RX-DMFIT to examine the effect that varying these attributes of the astrophysical model has on the profile, spectrum, and total flux resulting from DM annihilation. Our results show that effects such as diffusion of charged particle byproducts can be ignored in the case of most large scale systems such as galaxy clusters, but can provide order of magnitude corrections in dwarfs and other galaxies under conservative assumptions for diffusion values. Additionally, we discussed the presence of X-ray radiation resulting from IC scattering of CMB photons as a secondary form of emission due to DM annihilation. We showed that the inclusion of diffusion effects can lead to relative increases in the X-ray band as relativistic electrons and positrons diffuse into regions of lower magnetic field, which can potentially provide new methods of searching for dark matter.

We used radio data of the Milky Way dSph Segue I to place constraints on the dark matter particle cross-section and find the best limits at low masses with $\tau^+\tau^-$ and $\mu^+\mu^-$ final states. The $\mu^+\mu^-$ channel in particular provides the most stringent constraints. Assuming a low diffusion value of $D_0 = 3 \times 10^{ 26}$ cm$^2$s$^{-1}$, this annihilation channel provides limits below the canonical thermal relic cross-section for masses below 100 GeV, with constraints roughly an order of magnitude greater at $M_{\chi} \approx 10$ GeV. However, when assuming the more conservative value for the diffusion constant of $D_0 = 3 \times 10^{ 26}$ cm$^2$s$^{-1}$ these constraints are diminished by a factor of $\sim 20 - 30$, demonstrating the impact of diffusion effects in smaller systems, and a need for a better understanding of diffusion in dwarfs. The constraints we found are competitive with previous analysis of dSphs using gamma-ray observations and, in the some cases such as the $\mu^+\mu^-$ final states with weak diffusion, considerably more stringent.

The RX-DMFIT tool offers a useful and versatile way to predict the synchrotron and inverse Compton emission from DM annihilations. This can aid in the design and planning of future observations by allowing the user to determine optimal observing frequencies and region sizes for dark matter searches. Also, the analysis performed by RX-DMFIT will be of great use in distinguishing astrophysical radio and X-ray signals from potential dark matter signals, particularly where diffusion effects have significant impact on the profile of emission due to dark matter annihilation. Radio and X-ray emission in astrophysical systems have the potential to provide highly competitive constraints on dark matter properties. Diffusion, magnetic field, and dark matter profile parameters all have significant impact on the expected radio and X-ray emission from dark matter annihilation, and better understanding of these features can greatly improve current constraints. 

\acknowledgments
This material is based upon work supported by the National Science Foundation under Grant No. 1517545. S.P. is partly supported by the US Department of Energy, grant number DE-SC0010107. E.S. is supported by the Netherlands Organization for Scientific Research (NWO) through a Vidi grant (PI: C. Weniger).


\bibliographystyle{JHEP}
\bibliography{ref}

\end{document}